\documentclass[nofootinbib]{revtex4}
\usepackage{graphicx}
\usepackage{dcolumn}
\usepackage{amsmath,amssymb,epsfig}
\usepackage{paralist}
\usepackage{comment}
\usepackage{graphicx}
\usepackage{multirow}
\usepackage{color,soul}
\usepackage{suffix}

\allowdisplaybreaks
\renewcommand{\vec}[1]{\boldsymbol{\mathrm{#1}}}

\begin{document}


\title{Wave-optical treatment of the shadow cast by a large sphere}

\author{Slava G. Turyshev$^{1}$, Viktor T. Toth$^2$
}

\affiliation{\vskip 3pt
$^1$Jet Propulsion Laboratory, California Institute of Technology,\\
4800 Oak Grove Drive, Pasadena, CA 91109-0899, USA
}%

\affiliation{\vskip 3pt
$^2$Ottawa, Ontario K1N 9H5, Canada
}%

\date{\today}

\begin{abstract}

We study the electromagnetic (EM) field in the shadow cast by a large opaque sphere. For this, we consider the scattering of a high frequency  monochromatic EM wave by the large sphere and develop a Mie theory that accounts for the presence of this obscuration. Applying fully absorbing boundary conditions, we  find a solution for the Debye potentials, which we use to determine the EM field in the shadow in the wave zone at large distances from the sphere. We use the standard tools available from the nuclear scattering theory to develop the wave-optical  treatment of the problem. Based on this treatment, we demonstrate that there is no EM field deep in the shadow, except for the field that is diffracted into the shadow by the edges of the sphere, as anticipated.

\end{abstract}


\maketitle

\section{Introduction}
\label{sec:intro}

Recent efforts to investigate the optical properties of the solar gravitational lens (SGL) \cite{Turyshev:2017,Turyshev-Toth:2017} led us to the realization that it is important to develop a wave-theoretical description of the electromagnetic (EM) field in the shadow of a large, opaque sphere, in a sufficiently flexible manner such that the discussion can later be extended to include optical contributions from the gravitational field, surrounding diffractive atmosphere (the solar corona) and other effects.

We are interested in determining the EM field behind a large, fully absorbing sphere, the diameter of which, $R_0$, is much larger than the wavelength $\lambda$ of the incident EM radiation, i.e., $R_0\gg\lambda$. Additionally, we consider the propagation of a high frequency EM wave, so that $k R_0\gg1$, where $k=2\pi/\lambda$ is the wavenumber. As a specific example, the radius of the Sun is $R_0\simeq 695,700$~km and the wavelength of interest is $\lambda\sim 1~\mu$m, thus both conditions are satisfied. Therefore, in some sense, the availability of these relationships and related small parameters should make the problem easily solvable.  Indeed, there are many solutions that deal with similar problems, but they generally rely on the geometric optics approximation. For instance, similar discussions exist in the literature related to optical scattering \cite{Kerker-book:1969,Born-Wolf:1999,vandeHulst-book-1981,Grandy-book-2000}, and also in the eikonal or high energy approximations \cite{Sharma-etal:1988,Sharma-Somerford:1990,Sharma-Sommerford-book:2006} of light or nuclear scattering \cite{Akhiezer-Pomeranchuk:1950,Glauber-Matthiae:1970,Semon-Taylor:1977,Friedrich-book-2013,Friedrich-book-2006}, but they were not fully suitable for our purposes, as most of them use scalar diffraction theory and geometric optics.

Our objective is to develop a wave-optical treatment of the shadow behind a large, opaque sphere.  The solution of this problem is not really intuitive as, in the case of the SGL, we are dealing with the very large dimensions of the Sun itself and significant heliocentric distances where the Sun forms the focal area (i.e., beyond 547 astronomical units). Yet we are interested in the coherent addition of the EM fields represented by rays enveloping the Sun with different impact parameters but with optical paths that are equal within a fraction of a wavelength. The peak of the point spread function that characterizes the beam of extreme intensity occupies the region around the optical axis with a radius of $\sim 10$~cm \cite{Turyshev-Toth:2017}. Solutions to this particular problem with such a hugely mismatched parameter set do not exist in the literature. Relying, in part, on methods developed in the existing literature and extending our own work, we develop, from first principles, a Mie theory that accounts for the shadow in terms of an infinite series representation of incident and scattered EM waves. We demonstrate that in the shadow region, the EM field vanishes as expected.

Our discussion begins in Sec.~\ref{sec:debye}, where we introduce our notation and present Maxwell's vacuum field equations in terms of Debye potentials. In Sec.~\ref{sec:mie}, we develop a Mie theory for the scattering. We introduce the fully absorbing boundary conditions and represent the EM field as a sum of incident and scattered waves.  We rely on the properties of the Riccati-Bessel functions and asymptotic expansions of the Legendre-polynomials to demonstrate that the EM field is indeed absent in the shadow region. Finally, a summary of our results is presented in Sec.~\ref{sec:end}.

\section{Electromagnetic field in terms of the Debye potentials}
\label{sec:debye}

We wish to describe light propagation in the vacuum, in the vicinity of a large, electrically neutral, opaque sphere. To this end, we use the source-free Maxwell equations:
{}
\begin{eqnarray}
{\rm rot}\,{\vec E}&=&-  \frac{1}{c}\frac{\partial \,{\vec H}}{\partial t},
\qquad ~{\rm div} \,{\vec E}\,=\,0,
\label{eq:rotE_fl}\\
{\rm rot}\,{\vec H}&=& \frac{1}{c}\frac{\partial \,{\vec E}}{\partial t},
\qquad \quad\,
{\rm div } \,{\vec B}\,=\,0.
\label{eq:rotH_fl}
\end{eqnarray}
Following closely the derivation presented in \cite{Born-Wolf:1999}, we now consider a solution to these equations. In the case of the sphere's static, spherically symmetric geometry, solving (\ref{eq:rotE_fl})--(\ref{eq:rotH_fl}) is quite straightforward. We obtain the complete solution of these  equations in spherical coordinates $(r,\theta,\phi)$ in terms of the electric and magnetic Debye potentials \cite{Born-Wolf:1999}, ${}^e\Pi$ and ${}^m\Pi$, as
{}
\begin{eqnarray}
{ { E}}_r&=&\frac{\partial^2 }{\partial r^2}
\big(r\,{}^e{\hskip -1pt}\Pi\big)+k^2 \big(r\,{}^e{\hskip -1pt}\Pi\big),
\qquad\qquad \qquad\qquad\,
{ {  H}}_r \,=\,
\frac{\partial^2}{\partial r^2}\big(r\,{}^m{\hskip -1pt}\Pi\big)+k^2\big(r\,{}^m{\hskip -1pt}\Pi\big),
\label{eq:Dr-em0}\\[3pt]
{ {  E}}_\theta&=&
\frac{1}{r}\frac{\partial^2 \big(r\,{}^e{\hskip -1pt}\Pi\big)}{\partial r\partial \theta}+\frac{ik}{r\sin\theta}
\frac{\partial\big(r\,{}^m{\hskip -1pt}\Pi\big)}{\partial \phi},
\qquad \qquad \quad
{ {  H}}_\theta\,=\,
-\frac{ik}{r\sin\theta} \frac{\partial\big(r\,{}^e{\hskip -1pt}\Pi\big)}{\partial \phi}+\frac{1}{r}\frac{\partial^2 \big(r\,{}^m{\hskip -1pt}\Pi\big)}{\partial r\partial \theta},
\label{eq:Dt-em0}\\[3pt]
{ {  E}}_\phi&=&
\frac{1}{r\sin\theta}
\frac{\partial^2 \big(r\,{}^e{\hskip -1pt}\Pi\big)}{\partial r\partial \phi}-\frac{ik}{r}
\frac{\partial\big(r\,{}^m{\hskip -1pt}\Pi\big)}{\partial \theta},
\qquad\qquad~~
{ { H}}_\phi\,=\,
\frac{ik}{r}\frac{\partial\big(r\,{}^e{\hskip -1pt}\Pi\big)}{\partial \theta}+\frac{1}{r\sin\theta}
\frac{\partial^2 \big(r\,{}^m{\hskip -1pt}\Pi\big)}{\partial r\partial \phi},
\label{eq:Dp-em0}
\end{eqnarray}
where $k=2\pi/\lambda$ is the wavenumber for the incident EM wave with wavelength $\lambda$. Both of the potentials ${}^e{\hskip -1pt}\Pi$ and  ${}^m{\hskip -1pt}\Pi$ satisfy the wave equation
\begin{eqnarray}
(\Delta+k^2)\,\Pi=0.
\label{eq:Pi-eq+wew1*+}
\end{eqnarray}

In spherical  coordinates, the solution to Eq.~(\ref{eq:Pi-eq+wew1*+}) is typically \cite{Born-Wolf:1999} obtained by separating variables:
\begin{eqnarray}
\Pi=\frac{1}{r}R(r)\Theta(\theta)\Phi(\phi).
\label{eq:Pi*}
\end{eqnarray}
In the following, we may also make use of the cylindrical $z$-coordinate, related to spherical coordinates by $z=r\cos\theta$.

Direct substitution into Eq.~(\ref{eq:Pi-eq+wew1*+}) reveals, after some algebra, that the functions $R$, $\Theta$ and $\Phi$ satisfy the following differential equations:
{}
\begin{eqnarray}
\frac{d^2 R}{d r^2}+\Big(k^2 -\frac{\alpha}{r^2}\Big)R&=&0,
\label{eq:R-bar*}\\
\frac{1}{\sin\theta}\frac{d}{d \theta}\Big(\sin\theta \frac{d \Theta}{d \theta}\Big)+\big(\alpha-\frac{\beta}{\sin^2\theta}\big)\Theta&=&0,
\label{eq:Th*}\\
\frac{d^2 \Phi}{d \phi^2}+\beta\Phi&=&0.
\label{eq:Ph*}
\end{eqnarray}

The solution to (\ref{eq:Ph*}) is given as usual \cite{Born-Wolf:1999}:
{}
\begin{eqnarray}
\Phi_m(\phi)=e^{\pm im\phi}  \quad\rightarrow \quad \Phi_m(\phi)=a_m\cos (m\phi) +b_m\sin (m\phi),
\label{eq:Ph_m}
\end{eqnarray}
with $\beta=m^2$,  and with $a_m$ and $b_m$ being integration constants.

Equation (\ref{eq:Th*}) is well known for spherical harmonics. Single-valued solutions to this equation exist when $\alpha=l(l+1)$ with $(l>|m|;l,m\in\mathbb{Z})$. With this condition, the solution to (\ref{eq:Th*}) becomes
{}
\begin{eqnarray}
\Theta_{lm}(\theta)&=&P^{(m)}_l(\cos\theta).
\label{eq:Th_lm}
\end{eqnarray}

Given these solutions, equation (\ref{eq:R-bar*}) for the radial function takes the form
{}
\begin{eqnarray}
\frac{d^2 R}{d r^2}+\Big(k^2-\frac{\ell(\ell+1)}{r^2}\Big)R&=&0.
\label{eq:R-bar-k*}
\end{eqnarray}

The general solution to this equation is well known  \cite{Born-Wolf:1999} and  may be given in terms of the Riccati-Bessel functions $\psi_\ell(kr)$ and $\chi_\ell(kr)$  (see discussion in Appendix \ref{sec:Riccati-B}) as
{}
\begin{eqnarray}
R&=&c_\ell \psi_\ell(kr)+ d_\ell\chi_\ell(kr),
\label{eq:R-s+}
\end{eqnarray}
where $c_\ell$ and $d_\ell$ are arbitrary constants.  As  the function $\psi_\ell(kr)$ is regular everywhere, including the origin and the function $\chi_\ell(kr)$ has a singularity at the origin, it is $\psi_\ell(kr)$ that is suitable to represent the field inside the sphere \cite{Born-Wolf:1999,Kerker-book:1969}. Therefore, to represent the EM field outside the sphere, we choose $c_\ell=1$ and $d_\ell=0$.

We adopt the geometry from  discussions of the Mie problem \cite{Born-Wolf:1999}.  Thus, the source is located at a large distance from the sphere along the $z$-axis, which goes through the center of the sphere and is parallel to the direction of propagation of the incident wave. The incident plane EM wave, $\propto e^{ikz}$, is emitted by a source located at large negative $z$-values and, after passing by the sphere, it propagates toward positive $z$-values.

\begin{figure}[t]
\includegraphics[width=0.6\linewidth]{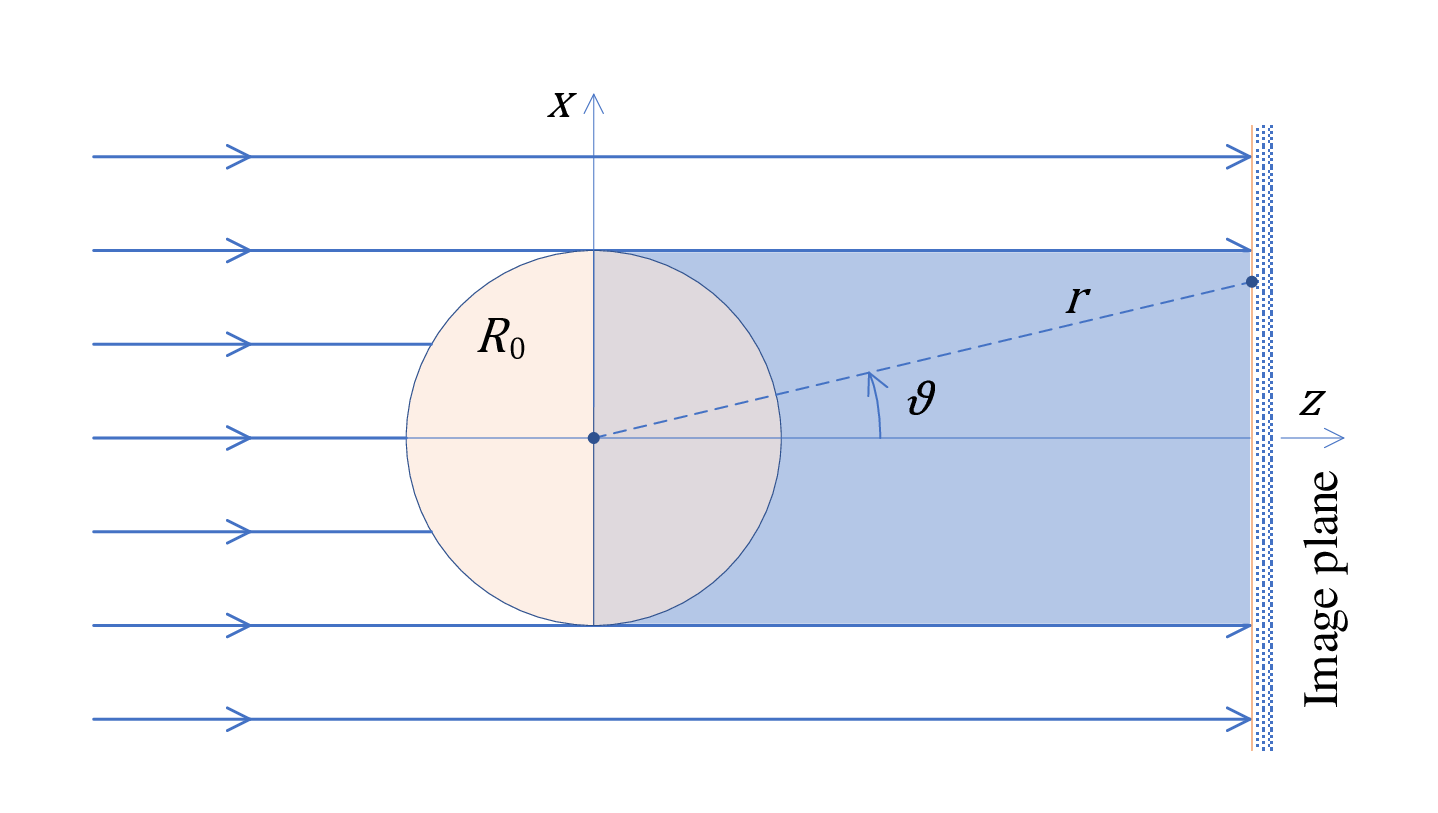}
\caption{
Geometry of the scattering problem. Body-centric spherical polar coordinate system $(r,\theta,\phi)$ (suppressed) as well as the $z$ and $x$ coordinates used to describe the diffraction of light on a large opaque sphere. \label{fig:geometry}}
\end{figure}

Next, we require our solution to (\ref{eq:Pi-eq+wew1*+}) to satisfy the asymptotic boundary condition at negative infinity by matching the incident plane EM wave $\propto e^{ikz}$. As a result, collecting (\ref{eq:Ph_m}), (\ref{eq:Th_lm}) and (\ref{eq:R-s+}), in the vacuum and because of the spherical symmetry of the problem, the solutions for the electric and magnetic potentials of the incident wave, ${}^e{\hskip -1pt}\Pi_0$ and ${}^m{\hskip -1pt}\Pi_0$, may be given in terms of a single potential $\Pi_0(r, \theta)$ (see  \cite{Born-Wolf:1999,Turyshev-Toth:2017} for details):
\begin{align}
  \left( \begin{aligned}
{}^e{\hskip -1pt}\Pi_0& \\
{}^m{\hskip -1pt}\Pi_0& \\
  \end{aligned} \right) =&  \left( \begin{aligned}
\cos\phi \\
\sin\phi  \\
  \end{aligned} \right) \,\Pi_0(r, \theta), & \hskip 2pt {\rm where}~~~~~
\Pi_0 (r, \theta)=
\frac{E_0}{k^2}\frac{1}{r}
\sum_{\ell=1}^\infty i^{\ell-1}\frac{2\ell+1}{\ell(\ell+1)}
\psi_\ell(kr) P^{(1)}_\ell(\cos\theta),
  \label{eq:Pi_ie*+*=}
\end{align}
where $E_0$ characterizes the energy density of the unperturbed EM wave at the source and $P_\ell^{(1)}$ are the associated Legendre-polynomials. To obtain the components of the EM field in the spherically symmetric and static geometry, we need to construct the following expressions (as was shown in  \cite{Turyshev-Toth:2017}):
\begin{eqnarray}
\alpha(r, \theta)&=&-\frac{1}{r^2}\frac{\partial}{\partial \theta}\Big[\frac{1}{\sin\theta} \frac{\partial}{\partial\theta}\big[\sin\theta\,(r\,\Pi)\big]\Big]\equiv \frac{\partial^2 \big(r\,{\hskip -1pt}\Pi\big)}{\partial r^2}
+k^2\big(r\,{\hskip -1pt}\Pi\big),
\label{eq:alpha*}\\
\beta(r, \theta)&=&\frac{1}{r}
\frac{\partial^2 \big(r\,{\hskip -1pt}\Pi\big)}{\partial r\partial \theta}+\frac{ik\big(r\,{\hskip -1pt}\Pi\big)}{r\sin\theta},
\label{eq:beta*}\\[0pt]
\gamma(r, \theta)&=&-\frac{1}{r\sin\theta}
\frac{\partial \big(r\,{\hskip -1pt}\Pi\big)}{\partial r}-\frac{ik}{r}
\frac{\partial\big(r\,{\hskip -1pt}\Pi\big)}{\partial \theta},
\label{eq:gamma*}
\end{eqnarray}
and insert them into
\begin{align}
  \left( \begin{aligned}
{  E}_r& \\
{  H}_r& \\
  \end{aligned} \right) =&  \left( \begin{aligned}
\cos\phi \\
\sin\phi  \\
  \end{aligned} \right) \,e^{-i\omega t}\alpha(r, \theta), &
    \left( \begin{aligned}
{  E}_\theta& \\
{  H}_\theta& \\
  \end{aligned} \right) =&  \left( \begin{aligned}
\cos\phi \\
\sin\phi  \\
  \end{aligned} \right) \,e^{-i\omega t}\beta(r, \theta), &
    \left( \begin{aligned}
{  E}_\phi& \\
{  H}_\phi& \\
  \end{aligned} \right) =&  \left( \begin{aligned}
\sin\phi \\
-\cos\phi  \\
  \end{aligned} \right) \,e^{-i\omega t}\gamma(r, \theta).
  \label{eq:DB-sol00*}
\end{align}

We will use these expression to study the EM field in the shadow produced by the large sphere.

\section{Diffraction of light by a large sphere}
\label{sec:mie}

\subsection{Fully absorbing boundary conditions}
\label{sec:bound-cond}

We consider parallel rays of light traveling in the $z$-direction, passing by a large sphere of radius $R_0\gg\lambda$. Each ray is characterized by its impact parameter $b$ with respect to the sphere. We consider an opaque sphere for which the rays with impact parameter $b\le R_0$ are completely absorbed by the sphere. In other words, we have the situation where all the radiation intercepted by the body is fully absorbed by it and no reflection or coherent reemission occurs. All intercepted radiation will be transformed into some other forms of energy, notably heat.

The observation above allows us to formally introduce the fully absorbing boundary conditions. To do this, we turn our attention to (\ref{eq:Pi_ie*+*=}) and remember that it was obtained by solving  the wave equation (\ref{eq:Pi-eq+wew1*+}). In quantum mechanics (\ref{eq:Pi-eq+wew1*+}) is identical to the time-independent Schr\"odinger equation \cite{Messiah:1968,Landau-Lifshitz:1989} of a free particle. Thus, the index of summation  $\ell$ in (\ref{eq:Pi_ie*+*=}) represents the quantum mechanical momentum for the $\ell$-th partial wave. In classical scattering theory \cite{Landau-Lifshitz:1989,Sharma-Sommerford-book:2006,Friedrich-book-2013}, the impact parameter $b$ is related to the quantum mechanical partial wave $\ell$ as $k \,b =\sqrt{\ell(\ell+1)}\approx \ell+{\textstyle\frac{1}{2}}$. This semiclassical analogy between $b$ and $\ell$  is useful to introduce our boundary conditions. For this, we identically rewrite $\Pi_0 (r, \theta)$ from (\ref{eq:Pi_ie*+*=}) as
{}
\begin{eqnarray}
\Pi_0 (r, \theta)&=&
\frac{E_0}{2k^2}\frac{1}{r}
\sum_{\ell=1}^\infty i^{\ell-1}\frac{2\ell+1}{\ell(\ell+1)} \Big\{
\zeta^{(+)}_\ell(kr)+\zeta^{(-)}_\ell(kr)\Big\} P^{(1)}_\ell(\cos\theta),
  \label{eq:Pi_g+p1}
\end{eqnarray}
where $\zeta_\ell^{(\pm)}$ are related to the Riccati-Bessel functions  by $\zeta_\ell^{(\pm)}(z)=\psi_\ell(z)\mp i\chi_\ell(z)$, given by (\ref{eq:spher-Hank1})--(\ref{eq:spher-Hank2}). The functions $\zeta_\ell^{(+)}$ and $\zeta_\ell^{(-)}$  correspond to radially incoming ($\propto e^{ikr}$) and radially outgoing ($\propto e^{-ikr}$) EM waves, respectively.

To implement the fully absorbing boundary conditions, it is convenient to introduce the image plane that is positioned at a large, positive-$z$ distance from the sphere and to consider the EM field in this plane (see Fig.~\ref{fig:geometry}). Finally, we require that no radially incoming EM waves reach the image plane for rays whose impact parameters $b\le R_0$ or, equivalently, for partial momenta $\ell \leq kR_0$. We implement these fully absorbing boundary conditions  by subtracting the radially incoming waves for $\ell \leq kR_0$ from the incident wave (as was discussed in \cite{Turyshev-Toth:2017}), which results in
{}
\begin{eqnarray}
\Pi(r, \theta)&=&
\frac{E_0}{2k^2}\frac{1}{r}
\sum_{\ell=1}^\infty i^{\ell-1}\frac{2\ell+1}{\ell(\ell+1)}\Big\{\zeta^{(+)}_\ell(kr)+\zeta^{(-)}_\ell(kr)\Big\} P^{(1)}_\ell(\cos\theta) -
\frac{E_0}{2k^2}\frac{1}{r}
\sum_{\ell=1}^{kR_0} i^{\ell-1}\frac{2\ell+1}{\ell(\ell+1)}\zeta^{(+)}_\ell(kr) P^{(1)}_\ell(\cos\theta) =\nonumber\\
&=&
\frac{E_0}{k^2}\frac{1}{r}
\sum_{\ell=1}^\infty i^{\ell-1}\frac{2\ell+1}{\ell(\ell+1)}\psi_\ell(kr) P^{(1)}_\ell(\cos\theta)-
\frac{E_0}{2k^2}\frac{1}{r}
\sum_{\ell=1}^{kR_0} i^{\ell-1}\frac{2\ell+1}{\ell(\ell+1)}\zeta^{(+)}_\ell(kr) P^{(1)}_\ell(\cos\theta).
\label{eq:Pi_g+p22}
\end{eqnarray}

Consider the field at a large distance $r\gg R_0$ from the sphere.
In this case, for $kr\gg \sqrt{\ell(\ell+1)}$, the asymptotic behavior of the function $\zeta^{(+)}_\ell(kr)$ from (\ref{eq:Pi_g+p22})  (as was shown in \cite{Turyshev-Toth:2017} and directly calculated by (\ref{eq:spher-Hank1}))  is given as
\begin{eqnarray}
\lim_{kr\rightarrow\infty} \zeta^{(+)}_\ell(kr)&=&\exp\Big[i\Big(kr-\frac{\pi}{2}(\ell+1)+\frac{\ell(\ell+1)}{2kr}\Big)\Big].
\label{eq:Fass}
\end{eqnarray}
Using (\ref{eq:Fass}) to represent the asymptotic behavior of $\zeta^{(+)}_\ell(kr)$, we present (\ref{eq:Pi_g+p22}) as
{}
\begin{eqnarray}
\Pi (r, \theta)&=&\Pi_0 (r, \theta)+\frac{e^{ikr}}{r}\frac{E_0}{2k^2}\sum_{\ell=1}^{kR_0} \frac{2\ell+1}{\ell(\ell+1)}e^{i\frac{\ell(\ell+1)}{2kr}}P^{(1)}_\ell(\cos\theta).
  \label{eq:Pi_g+p0}
\end{eqnarray}

The first term in (\ref{eq:Pi_g+p0}) is the Debye potential representing the incident plane wave \cite{Born-Wolf:1999} (this solution may be derived from the solution obtained in \cite{Turyshev-Toth:2017}). The second term is responsible for the geometric shadow behind the body.

Introducing the obscuration amplitude,
{}
\begin{eqnarray}
  f_{\tt ob}(\theta)&=&\frac{E_0}{2k^2}\sum_{\ell=1}^{kR_0} \frac{2\ell+1}{\ell(\ell+1)}e^{i\frac{\ell(\ell+1)}{2kr}}
P^{(1)}_\ell(\cos\theta),
  \label{eq:Pi_g+p22+f}
\end{eqnarray}
allows us to present the Debye potential in the following compact form:
{}
\begin{eqnarray}
\Pi(r, \theta)&=&\Pi_0 (r, \theta)+\Pi_{\tt ob} (r, \theta)=\Pi_0 (r, \theta)+f_{\tt ob}(\theta)\frac{e^{ikr}}{r},
  \label{eq:Pi-g+p}
\end{eqnarray}
where $\Pi_{\tt ob} (r, \theta)$ is the Debye potential representing a fictitious the EM field due to the obscuration. This fictitious EM field is a very useful concept. As shown below, it allows the application of well developed methods of nuclear scattering theory in describing the  spherical obscuration. This field is induced by the material within the sphere to precisely match the incident EM field, so that the total EM field on the image plane vanishes.

Eq.~(\ref{eq:Pi-g+p}) is our main result, as it contains all information needed to describe propagation of monochromatic EM waves in the vicinity of a large opaque sphere.

\subsection{Debye potential in the shadow}
\label{sec:Deb-pot}

To evaluate $f_{\tt ob}(\theta)$ from (\ref{eq:Pi_g+p22+f}), we use the asymptotic representation for $P^{(1)}_l(\cos\theta)$ from \cite{Korn-Korn:1968}, valid when $\ell\to\infty$:
{}
\begin{align}
P^{(1)}_\ell(\cos\theta)  &=
\dfrac{-\ell}{\sqrt{2\pi \ell \sin\theta}}\Big(e^{i(\ell+\frac{1}{2})\theta+i\frac{\pi}{4}}+e^{-i(\ell+\frac{1}{2})\theta-i\frac{\pi}{4}}\Big)+{\cal O}(\ell^{-\textstyle\frac{3}{2}}) ~~~~~\textrm{for}~~~~~ 0<\theta<\pi.
\label{eq:P1l<}
\end{align}

This approximation can be used to transform (\ref{eq:Pi_g+p22+f}) as
{}
\begin{eqnarray}
f_{\tt ob}(\theta)&=&
\frac{E_0}{2k^2}\sum_{\ell=1}^{kR_0} \frac{2\ell+1}{\ell(\ell+1)}\frac{(-\ell)}{\sqrt{2\pi \ell \sin\theta}}\,e^{i\frac{\ell(\ell+1)}{2kr}}\Big(e^{i(\ell+\frac{1}{2})\theta+i\frac{\pi}{4}}+e^{-i(\ell+\frac{1}{2})\theta-i\frac{\pi}{4}}\Big).
\label{eq:P_sum}
\end{eqnarray}

At this point, we may replace the sum in (\ref{eq:P_sum}) with an integral:
{}
\begin{eqnarray}
f_{\tt ob}(\theta)&=&
\frac{E_0}{2k^2}\int_{1}^{kR_0} \frac{2\ell+1}{\ell(\ell+1)}\frac{(-\ell) d\ell}{\sqrt{2\pi \ell \sin\theta}}\,e^{i\frac{\ell(\ell+1)}{2kr}}\Big(e^{i(\ell+\frac{1}{2})\theta+i\frac{\pi}{4}}+e^{-i(\ell+\frac{1}{2})\theta-i\frac{\pi}{4}}\Big),~~~~~~
\label{eq:Pi_s_exp1}
\end{eqnarray}
and evaluate this integral by the method of stationary phase. This method allows us to evaluate integrals of the type
{}
\begin{equation}
I=\int A(\ell)e^{i\varphi(\ell)}d\ell, \qquad
\ell\in\mathbb{R},
\label{eq:stp-1}
\end{equation}
where the amplitude $A(\ell)$ is a slowly varying function of $\ell$, while $\varphi(\ell)$ is a rapidly varying function of $\ell$.
The integral (\ref{eq:stp-1}) may be replaced, to good approximation, with a sum over the points of stationary phase, $\ell_0\in\{\ell_{1,2,..}\}$, for which $d\varphi/d\ell=0$ (and defining $\varphi''=d^2\varphi/d\ell^2$):
{}
\begin{equation}
I\simeq\sum_{\ell_0\in\{\ell_{1,2,..}\}} A(\ell_0)e^{i\varphi(\ell_0)}e^{i{\textstyle\frac{\pi}{4}}}\sqrt{\frac{2\pi}{\varphi''(\ell_0)}}.
\label{eq:stp-2}
\end{equation}

The $\ell$-dependent part of the phase of (\ref{eq:Pi_s_exp1}) is of the form
{}
\begin{equation}
\varphi_{\pm}(\ell)=\pm\big((\ell+\textstyle{\frac{1}{2}})\theta+\textstyle{\frac{\pi}{4}}\big)+\dfrac{\ell(\ell+1)}{2kr}.
\label{eq:S-l}
\end{equation}
The phase is stationary when $d\varphi_{\pm}/d\ell=0$, which implies
{}
\begin{equation}
\pm\theta= -\frac{\ell+{\textstyle\frac{1}{2}}}{kr}.
\label{eq:S-l-pri}
\end{equation}
In the semiclassical approximation of the theory of particle scattering (for discussion, see \cite{Sharma-Sommerford-book:2006,Friedrich-book-2013}), $\ell$ represents partial momenta, related to the impact parameter and the wavenumber by
{}
\begin{equation}
\ell+{\textstyle\frac{1}{2}}= kb.
\label{eq:S-l-pri-p}
\end{equation}
For small angles $\theta$ (or, large distances from the sphere, $R_0\ll r$),  Eq.~(\ref{eq:S-l-pri}) yields $\pm\sin\theta=-b/r$.  As a result, we see that the points of stationary phase satisfy the equation
{}
\begin{equation}
b=r\cos(\theta\pm\textstyle{\frac{\pi}{2}}),
\label{eq:theta-b0}
\end{equation}
which is the equation for a family of parallel straight lines.

The largest value of the impact parameter for (\ref{eq:Pi_s_exp1}) is $b=R_0$. The two lines given by $R_0=r\cos(\theta\mp\textstyle{\frac{\pi}{2}})$ represent the boundary that coincides with two rays of light that are just grazing the sphere on opposite sides while traveling the forward direction, $0\leq\theta\leq \frac{\pi}{2}$, setting the boundary of the geometric shadow behind the sphere.

Let us demonstrate that there is indeed no EM field in the shadow behind the sphere. For this, we continue to evaluate the integral in (\ref{eq:Pi_s_exp1}) using the method of stationary phase. From  (\ref{eq:S-l-pri}), we determine
{}
\begin{equation}
\ell_0+{\textstyle\frac{1}{2}} = \mp kr \theta,
\label{eq:ell0}
\end{equation}
which allows us to compute
{}
\begin{equation}
\varphi_{\pm}(\ell_0)=\pm\textstyle{\frac{\pi}{4}}-\textstyle{\frac{1}{2}}\theta^2 kr,
\qquad \varphi''_{\pm}(\ell_0)=\dfrac{1}{kr}.
\label{eq:S-l2}
\end{equation}
As a result, for each of the two areas with respect to the $z$-axis---above it given by ``$+$''-sign and below, given by ``$-$''-sign---we obtain
{}
\begin{eqnarray}
f_{\tt ob}(\theta)&=&
\frac{E_0}{2k^2} \frac{-2}{\sqrt{\mp1}  \sin\theta}e^{i\big(\pm\textstyle{\frac{\pi}{4}}-\textstyle{\frac{1}{2}}\theta^2 kr\big)+i\textstyle{\frac{\pi}{4}}}=-\frac{E_0}{k^2\sin\theta}e^{-ikr\textstyle{\frac{1}{2}}\theta^2},
\label{eq:Pi_s_exp4+=}
\end{eqnarray}
which results in the following expression for the Debye potential of the scattered wave:
{}
\begin{eqnarray}
\Pi_{\tt ob}(r,\theta)&=& \frac{e^{ikr}}{r}f_{\tt ob}(\theta)=-\frac{1}{k^2r\sin\theta}e^{ikr\big(1-\textstyle{\frac{1}{2}}\theta^2\big)}=-\frac{1}{k^2 r\sin\theta}e^{ikr\cos\theta}.
\label{eq:Pi_s_exp4+}
\end{eqnarray}

Putting this result into (\ref{eq:Pi-g+p}), we have
\begin{eqnarray}
\Pi (r, \theta)&=&
\Pi_0 (r, \theta)-\frac{1}{k^2r \sin\theta}e^{ikr\cos\theta}.
\label{eq:P_tot2}
\end{eqnarray}

In Eq.~(\ref{eq:Pi_ie*+*=}), $\Pi_0(r,\theta)$ was presented in the form of an infinite sum. To evaluate $\Pi_0(r,\theta)$ further, it is helpful to obtain a closed form expression. To this purpose, following the method presented in \cite{Turyshev-Toth:2017}, we use the $E_r$ component of the incident plane EM wave (see \cite{Born-Wolf:1999}):
{}
\begin{eqnarray}
E_r&=&
-E_0\frac{\cos\phi}{ikr}
\frac{\partial \psi_0(r,\theta)}{\partial\theta}e^{-i\omega t},
\label{eq:Dr-em"}
\end{eqnarray}
where $\psi_0(r,\theta)=E_0e^{ikz}$ is the incident scalar wave  (see details in \cite{Turyshev-Toth:2017}). Equating this expression with the expression for $E_r$ in (\ref{eq:DB-sol00*}), we obtain, for the incident wave,
{}
\begin{eqnarray}
E_{r}&=&-e^{-i\omega t}\,\frac{\cos \phi}{r^2}\frac{\partial}{\partial \theta}\Big[\frac{1}{\sin\theta} \frac{\partial}{\partial\theta}\big[\sin\theta\,(r\,\Pi_0)\big]\Big]=-e^{-i\omega t}\,\frac{\cos\phi}{ikr}\frac{\partial \psi_0}{\partial\theta}.
\label{eq:vec_D_r*++}
\end{eqnarray}
{}
This result yields the following equation for the incident wave Debye potential $\Pi_0$:
{}
\begin{eqnarray}
\frac{\partial}{\partial \theta}\Big[\frac{1}{\sin\theta} \frac{\partial}{\partial\theta}\big[\sin\theta\,\Pi_0\big]\Big]=-\frac{i}{k}\frac{\partial \psi_0}{\partial\theta}.
\label{eq:vec_D_r*+}
\end{eqnarray}
We may now integrate this equation with respect to $\theta$ to obtain
{}
\begin{eqnarray}
\frac{\partial}{\partial\theta}\big[\sin\theta\,\Pi_0\big]=-\frac{i}{k}\sin\theta\big[\psi_0(r,\theta)+c(r)\big],
\label{eq:vec_D_r1*0}
\end{eqnarray}
where $c(r)$ is constant with respect to the integration variable $\theta$. Integrating again from $\pi$ to $\theta$, we have
{}
\begin{eqnarray}
\Pi_0(r,\theta)=-\frac{i}{k\sin\theta}\int_\pi^\theta\big[\psi_0(r,\theta')+c(r)\big]\sin\theta' d\theta'.
\label{eq:vec_D_r1*}
\end{eqnarray}
Given $\psi_0(r,\theta)=E_0e^{ikz}$, we can evaluate the integral (\ref{eq:vec_D_r1*}) as
{}
\begin{eqnarray}
\Pi_0(r,\theta)&=&\frac{E_0}{k^2r\sin\theta}\Big(e^{ikr\cos\theta}-e^{-ikr}+c(r)(\cos\theta+1)\Big).
\label{eq:sol-Pi0*}
\end{eqnarray}
Using $c(r)=-\frac{1}{2}(e^{ikr}-e^{-ikr})$, we satisfy the requirement for $\Pi_0$ to be finite for any $\theta$ and obtain the following expression for the Debye potential:
{}
\begin{eqnarray}
\Pi_0 (r, \theta)&=&
\frac{1}{k^2r \sin\theta}\Big(e^{ikr\cos\theta}-e^{ikr}+{\textstyle\frac{1}{2}}(1-\cos\theta)\big(e^{ikr}-e^{-ikr}\big)\Big).
\label{eq:P0}
\end{eqnarray}
(Note that an identical expression may be obtained from the solution found in \cite{Turyshev-Toth:2017} by taking the limit $r_g\rightarrow 0$.)

Substituting this expression into (\ref{eq:P_tot2}), we obtain the Debye potential in the shadow behind the sphere:
\begin{align}
  \left( \begin{aligned}
{}^e{\hskip -1pt}\Pi& \\
{}^m{\hskip -1pt}\Pi& \\
  \end{aligned} \right) =&  \left( \begin{aligned}
\cos\phi \\
\sin\phi  \\
  \end{aligned} \right) \,\Pi(r, \theta), & \hskip -20pt {\rm where}  \qquad
\Pi (r, \theta)=-
\frac{1}{2k^2r \sin\theta}\Big((1+\cos\theta)e^{ikr}+(1-\cos\theta)e^{-ikr}\Big).
\label{eq:P-scat}
\end{align}
Using this result in (\ref{eq:alpha*})--(\ref{eq:gamma*}), we confirm that these expressions all vanish in the shadow:
{}
\begin{eqnarray}
\alpha(r, \theta)&=&\frac{1}{2k^2r^2}\frac{\partial}{\partial \theta}\Big[\frac{1}{\sin\theta} \frac{\partial}{\partial\theta}\Big((1+\cos\theta)e^{ikr}+(1-\cos\theta)e^{-ikr}\Big)\Big]=0,
\label{eq:alpha*1}\\
\beta(r, \theta)&=&-\frac{ik}{2k^2r}\Big\{
\frac{\partial }{\partial \theta}\Big[
\frac{(1+\cos\theta)e^{ikr}-(1-\cos\theta)e^{-ikr}}{\sin\theta}\Big]+\frac{(1+\cos\theta)e^{ikr}+(1-\cos\theta)e^{-ikr}}{\sin^2\theta}\Big\}=0,
\label{eq:beta*1}\\[0pt]
\gamma(r, \theta)&=&\frac{ik}{2k^2r}\Big\{\frac{(1+\cos\theta)e^{ikr}-(1-\cos\theta)e^{-ikr}}{\sin^2\theta}+
\frac{\partial }{\partial \theta}\Big[
\frac{(1+\cos\theta)e^{ikr}+(1-\cos\theta)e^{-ikr}}{\sin\theta}\Big]\Big\}=0.
\label{eq:gamma*1}
\end{eqnarray}
As $\alpha=\beta=\gamma=0$,  according to (\ref{eq:DB-sol00*}), the EM field is absent: there is no light  in the shadow.

At the same time, for impact parameters $b>R_0$, the EM field is given fully by the incident wave, with the Debye potential $\Pi_0(r,\theta)$ from (\ref{eq:Pi_ie*+*=}) or, equivalently, from (\ref{eq:P0}). This field will diffract into the shadow with usual diffraction-limited divergence angle of $\sim \lambda/D_0$, where $D_0$ is the diameter of the spherical obscuration.  

\subsection{EM field in the shadow}
\label{sec:EM-field-near-axis}

To verify the results obtained above, we compute the fictitious EM field produced by the obscuration. The corresponding Debye potentials for the obscuration field  take the form
{}
\begin{align}
  \left( \begin{aligned}
{}^e{\hskip -1pt}\Pi_{\tt ob}& \\
{}^m{\hskip -1pt}\Pi_{\tt ob}& \\
  \end{aligned} \right) =&  \left( \begin{aligned}
\cos\phi \\
\sin\phi  \\
  \end{aligned} \right) \,\Pi_{\tt ob}(r, \theta), \qquad{\rm where}\qquad
  \Pi_{\tt ob} (r, \theta)=f_{\tt ob}(\theta)\frac{e^{ikr}}{r},
  \label{eq:Pi_ie*+8p}
\end{align}
with $f_{\tt ob}(\theta)$ given by (\ref{eq:Pi_g+p22+f}).
We will use (\ref{eq:Pi_ie*+8p}) to derive the components of the EM field produced by the scattered wave in the wave zone.
Indeed, using (\ref{eq:alpha*})--(\ref{eq:gamma*}) for the components of the EM field (\ref{eq:DB-sol00*}), we have ${    E}_r={    H}_r={\cal O}(r^{-2})$ and the rest of the components are given as
{}
\begin{eqnarray}
\left( \begin{aligned}
{    E}_\theta& \\
{    H}_\theta& \\
  \end{aligned} \right)_{\tt ob} &=&  ik\frac{e^{ikr}}{r}\Big\{\frac{\partial f_{\tt ob}(\theta)}{\partial \theta}+\frac{f_{\tt ob}(\theta)}{\sin\theta}\Big\}
\left( \begin{aligned}
\cos\phi \\
\sin\phi  \\
  \end{aligned} \right) \,e^{-i\omega t},
  \label{eq:DB-th}\\
  \left( \begin{aligned}
{    E}_\phi& \\
{    H}_\phi& \\
  \end{aligned} \right)_{\tt ob} &=&  -ik\frac{e^{ikr}}{r}\Big\{\frac{\partial f_{\tt ob}(\theta)}{\partial \theta}+\frac{f_{\tt ob}(\theta)}{\sin\theta}\Big\}
\left( \begin{aligned}
\sin\phi \\
-\cos\phi  \\
  \end{aligned} \right) \,e^{-i\omega t}.
  \label{eq:DB-ph}
\end{eqnarray}

Using (\ref{eq:Pi_g+p22+f}), we may rewrite the subexpression in curly braces as follows:
\begin{eqnarray}
\frac{\partial f_{\tt ob}(\theta)}{\partial \theta}+\frac{f_{\tt ob}(\theta)}{\sin\theta}=\frac{E_0}{2k^2}\sum_{\ell=1}^{kR_0} \frac{2\ell+1}{\ell(\ell+1)}
e^{i\frac{\ell(\ell+1)}{2kr}}
\Big\{\frac{\partial P^{(1)}_\ell(\cos\theta) }{\partial \theta}+\frac{P^{(1)}_\ell(\cos\theta) }{\sin\theta}\Big\}.
  \label{eq:f_vp}
\end{eqnarray}
{}
Introducing the obscuration amplitude function
\begin{eqnarray}
S(\theta)&=& \frac{1}{2} \sum_{\ell=1}^{kR_0} \frac{2\ell+1}{\ell(\ell+1)}e^{i\frac{\ell(\ell+1)}{2kr}}
\Big\{\frac{\partial P^{(1)}_\ell(\cos\theta) }{\partial \theta}+\frac{P^{(1)}_\ell(\cos\theta) }{\sin\theta}
\Big\},
\label{eq:S1-v0p}
\end{eqnarray}
we can present (\ref{eq:DB-th})--(\ref{eq:DB-ph}) as
{}
\begin{eqnarray}
  \left( \begin{aligned}
{    E}_\theta& \\
{    H}_\theta& \\
  \end{aligned} \right)_{\tt ob} &=& -E_0\left( \begin{aligned}
\cos\phi \\
\sin\phi  \\
  \end{aligned} \right)\frac{e^{i(kr-\omega t)}}{ikr}S(\theta)
,\quad
\left( \begin{aligned}
{    E}_\phi& \\
{    H}_\phi& \\
  \end{aligned} \right)_{\tt ob} =E_0\left( \begin{aligned}
\sin\phi \\
-\cos\phi  \\
  \end{aligned} \right)\frac{e^{i(kr-\omega t)}}{ikr}S(\theta).
  \label{eq:DB-t-pl}
\end{eqnarray}

To evaluate the magnitude of the amplitude function $S(\theta)$, we  need to establish the asymptotic behavior of the expressions involving Legendre-polynomials in (\ref{eq:S1-v0p}). There are exit two relevant expressions \cite{vandeHulst-book-1981}: one is  for fixed $w=(\ell+{\textstyle\frac{1}{2}})\theta$ and $\ell\rightarrow\infty$, which is given as:
{}
\begin{eqnarray}
\pi_\ell(\cos\theta)=&\dfrac{P^{(1)}_\ell(\cos\theta)}{\sin\theta}&= {\textstyle\frac{1}{2}}\ell(\ell+1)\Big(J_0(w)+J_2(w)\Big),
\label{eq:pi-l}\\
\tau_\ell(\cos\theta)=&\dfrac{dP^{(1)}_\ell(\cos\theta)}{d\theta}&= {\textstyle\frac{1}{2}}\ell(\ell+1)\Big(J_0(w)-J_2(w)\Big).
\label{eq:tau-l}
\end{eqnarray}
The area with $\theta\approx0$ is in the region of the geometric shadow behind the sphere, these approximations are not very useful. However, there exists another form for the expressions for the asymptotic behavior of $\pi_\ell(\theta)$ and $\tau_\ell(\theta)$. For fixed $\theta$ and $\ell\rightarrow\infty$ we have \cite{vandeHulst-book-1981}
{}
\begin{eqnarray}
\pi_\ell(\cos\theta)=&\dfrac{P^{(1)}_\ell(\cos\theta)}{\sin\theta}
&= \Big(\frac{2\ell}{\pi\sin^3\theta}\Big)^{\frac{1}{2}} \sin\Big((\ell+{\textstyle\frac{1}{2}})\theta-{\textstyle\frac{\pi}{4}}\Big),
\label{eq:pi-l*}\\
\tau_\ell(\cos\theta)=&\dfrac{dP^{(1)}_\ell(\cos\theta)}{d\theta}
&=  \Big(\frac{2\ell^3}{\pi\sin\theta}\Big)^{\frac{1}{2}} \cos\Big((\ell+{\textstyle\frac{1}{2}})\theta-{\textstyle\frac{\pi}{4}}\Big).
\label{eq:tau-l*}
\end{eqnarray}

For any large $\ell$, formulae (\ref{eq:pi-l*})--(\ref{eq:tau-l*}) are insufficient in a region close to the forward direction $(\theta=0$). In the forward region they are complemented by the asymptotic formulae (\ref{eq:pi-l})--(\ref{eq:tau-l}). More precisely, the formulae (\ref{eq:pi-l*})--(\ref{eq:tau-l*}) hold for $\sin\theta\gg1/\ell$ and those given by (\ref{eq:pi-l})--(\ref{eq:tau-l}) hold for $\theta\ll1$. The overlapping domain is $1/\ell\ll\sin\theta\ll 1$.

Considering (\ref{eq:pi-l*}) and (\ref{eq:tau-l*}), we see that $\tau_\ell$ is $\ell\sin\theta$ times larger than of $\pi_\ell$;
for most angles except the near forward direction, this represents a difference of an order of magnitude or more. Thus, we may neglect  the contribution of $\pi_\ell$ and use only $\tau_\ell$. With these approximations, the amplitude $S(\theta)$, defined by (\ref{eq:S1-v0p}), takes the following form in the region given by $\sin\theta\gg 1/\ell$:
{}
\begin{eqnarray}
S(\theta)&=& \frac{2}{\sqrt{2\pi\sin\theta}}
\sum_{\ell=1}^{kR_0} e^{i\frac{\ell(\ell+1)}{2kr}}\sqrt{\ell}
\cos\Big((\ell+{\textstyle\frac{1}{2}})\theta-{\textstyle\frac{\pi}{4}}\Big).
\label{eq:S1-v0s}
\end{eqnarray}
which we evaluate using the method of stationary phase. Representing (\ref{eq:S1-v0s}) in the form of an integral over $\ell$, we have
{}
\begin{eqnarray}
S(\theta)&=& \frac{-i}{\sqrt{2\pi\sin\theta}}
\int_{\ell=1}^{kR_0} \sqrt{\ell} d\ell e^{i\frac{\ell(\ell+1)}{2kr}}\Big(e^{i(\ell+{\textstyle\frac{1}{2}})\theta+i{\textstyle\frac{\pi}{4}}}-e^{-i(\ell+{\textstyle\frac{1}{2}})\theta-i{\textstyle\frac{\pi}{4}}}\Big).
\label{eq:S1-v0s+int*}
\end{eqnarray}
and the $\ell$-dependent parts of the phase, $\varphi_{\pm}(\ell)$, is identical to (\ref{eq:S-l}), allowing us to borrow some of our earlier results. Using (\ref{eq:ell0})--(\ref{eq:S-l2}) as we apply (\ref{eq:stp-1})--(\ref{eq:stp-2}) to (\ref{eq:S1-v0s+int*}), we have
{}
\begin{equation}
A(\ell_0)=\sqrt{\ell_0}=\sqrt{\mp\theta kr}, \qquad
\sqrt{\frac{2\pi}{\varphi''(\ell_0)}}=\sqrt{2\pi kr}
\qquad \rightarrow
\qquad
A(\ell_0)\sqrt{\frac{2\pi}{\varphi''(\ell_0)}}\simeq \sqrt{\mp2\pi \theta} kr,
\label{eq:S-l3p0}
\end{equation}
with the ``$+$''-sign and ``$-$''-sign respectively representing areas above and below the $z$-axis. Therefore, the expression for the $S_{\pm}(\theta)$ takes the form
{}
\begin{eqnarray}
S_{\pm}(\theta)&=&
\frac{\mp i}{\sqrt{2\pi\sin\theta}}\sqrt{\mp2\pi \theta} kr e^{i\big(\pm\textstyle{\frac{\pi}{4}}-\textstyle{\frac{1}{2}}\theta^2 kr\big)+i\textstyle{\frac{\pi}{4}}}=
ikr e^{-i\textstyle{\frac{1}{2}}\theta^2 kr}.
\label{eq:Pi_s_exp4+1}
\end{eqnarray}

With this result, (\ref{eq:DB-t-pl}) becomes
{}
\begin{eqnarray}
  \left( \begin{aligned}
{  E}_\theta& \\
{  H}_\theta& \\
  \end{aligned} \right)_{\tt ob} &=&
 -E_0e^{ikr(1-{\textstyle{\frac{1}{2}}}\theta^2)}
\left( \begin{aligned}
\cos\phi \\
\sin\phi  \\
  \end{aligned} \right) \,e^{-i\omega t}=
  -E_0
\left( \begin{aligned}
\cos\phi \\
\sin\phi  \\
  \end{aligned} \right) \,e^{i(kr\cos\theta-\omega t)}+{\cal O}(\theta^3),
  \label{eq:DB-t-pl=p1}
\\
\left( \begin{aligned}
{  E}_\phi& \\
{  H}_\phi& \\
  \end{aligned} \right)_{\tt ob} &=&E_0e^{ikr(1-{\textstyle{\frac{1}{2}}}\theta^2)}
\left( \begin{aligned}
\sin\phi \\
-\cos\phi  \\
  \end{aligned} \right) \,e^{-i\omega t}=E_0
\left( \begin{aligned}
\sin\phi \\
-\cos\phi  \\
  \end{aligned} \right) \,e^{i(kr\cos\theta-\omega t)} +{\cal O}(\theta^3).
  \label{eq:DB-t-pl=p2}
\end{eqnarray}

At the same time, the non-zero components of the incident EM field due to the Debye potential $\Pi_0$ are given by
{}
\begin{eqnarray}
    \left( \begin{aligned}
{    E}_\theta& \\
{    H}_\theta& \\
  \end{aligned} \right)_0 = E_0
\left( \begin{aligned}
\cos\phi \\
\sin\phi  \\
  \end{aligned} \right) \,e^{i(kr\cos\theta-\omega t)}, \qquad \left( \begin{aligned}
{    E}_\phi& \\
{    H}_\phi& \\
  \end{aligned} \right)_0 =-E_0
  \left( \begin{aligned}
\sin\phi \\
-\cos\phi  \\
  \end{aligned} \right) \,e^{i(kr\cos\theta-\omega t)}.
  \label{eq:DB-t-pl=p202}
\end{eqnarray}
As a result the total EM field, given by the sum of (\ref{eq:DB-t-pl=p1})--(\ref{eq:DB-t-pl=p2}) and (\ref{eq:DB-t-pl=p202}), vanishes near the optical axis for $\sin\theta\gg\ell$.

\subsection{EM field on the optical axis}
\label{sec:EM-field-on-axis}

To compute the field exactly on the optical axis, for $\theta=0$, we use the expression for the total Debye potential behind the sphere, for impact parameters $b\leq R$, which may be derived from  (\ref{eq:Pi_g+p22}) as
{}
\begin{eqnarray}
\Pi^{(-)}_{\tt ob}(r, \theta)&=&\frac{E_0}{2k^2}\frac{1}{r}
\sum_{\ell=1}^{kR_0} i^{\ell-1}\frac{2\ell+1}{\ell(\ell+1)}\zeta^{(-)}_\ell(kr) P^{(1)}_\ell(\cos\theta).
\label{eq:Pi_g+p22=0}
\end{eqnarray}

Similarly to (\ref{eq:Fass}), we take the asymptotic behavior of the function $\zeta^{(-)}_\ell(kr)$ for $kr\rightarrow \infty$ from (\ref{eq:spher-Hank2})  as
\begin{eqnarray}
\lim_{kr\rightarrow\infty} \zeta^{(-)}_\ell(kr)&=&\exp\Big[-i\Big(kr-\frac{\pi}{2}(\ell+1)+\frac{\ell(\ell+1)}{2kr}\Big)\Big].
\label{eq:Fass-}
\end{eqnarray}
Using this expression, we transform (\ref{eq:Pi_g+p22=0}):
{}
\begin{eqnarray}
\Pi^{(-)}_{\tt ob}(r, \theta)&=&\frac{E_0}{2k^2}\frac{e^{-ikr}}{r}
\sum_{\ell=1}^{kR_0} (-1)^\ell\frac{2\ell+1}{\ell(\ell+1)}e^{-i\frac{\ell(\ell+1)}{2kr}} P^{(1)}_\ell(\cos\theta).
\label{eq:Pi_g+p22=0=}
\end{eqnarray}

The corresponding Debye potentials for the obscuration EM field in this case take the form
{}
\begin{align}
  \left( \begin{aligned}
{}^e{\hskip -1pt}\Pi^{(-)}_{\tt ob}& \\
{}^m{\hskip -1pt}\Pi^{(-)}_{\tt ob}& \\
  \end{aligned} \right) =&  \left( \begin{aligned}
\cos\phi \\
\sin\phi  \\
  \end{aligned} \right) \,\Pi^{(-)}_{\tt ob}(r, \theta), \qquad{\rm where}\qquad
  \Pi^{(-)}_{\tt ob} (r, \theta)=f^{(-)}_{\tt ob}(\theta)\frac{e^{-ikr}}{r},
  \label{eq:Pi_ie*+8p=}
\end{align}
with obscuration amplitude $f^{(-)}_{\tt ob}(\theta)$ having the form
{}
\begin{eqnarray}
f^{(-)}_{\tt ob}(\theta)&=&\frac{E_0}{2k^2}
\sum_{\ell=1}^{kR_0} (-1)^\ell\frac{2\ell+1}{\ell(\ell+1)}e^{-i\frac{\ell(\ell+1)}{2kr}} P^{(1)}_\ell(\cos\theta).
\label{eq:Pi_g+p22=0=f}
\end{eqnarray}
We use (\ref{eq:Pi_ie*+8p=}) and (\ref{eq:Pi_g+p22=0=f}) to derive the components of the EM field produced by the scattered wave in the wave zone.
Indeed, using (\ref{eq:alpha*})--(\ref{eq:gamma*}), for the components of the EM field (\ref{eq:DB-sol00*}), we have ${ E}^{(-)}_r={ H}^{(-)}_r={\cal O}(r^{-2})$ and the rest of the components are given as
{}
\begin{eqnarray}
\left( \begin{aligned}
{    E}^{(-)}_\theta& \\
{    H}^{(-)}_\theta& \\
  \end{aligned} \right)_{\tt ob} &=&  ik\frac{e^{-ikr}}{r}\Big\{-\frac{\partial f^{(-)}_{\tt ob}(\theta)}{\partial \theta}+\frac{f^{(-)}_{\tt ob}(\theta)}{\sin\theta}\Big\}
\left( \begin{aligned}
\cos\phi \\
\sin\phi  \\
  \end{aligned} \right) \,e^{-i\omega t},
  \label{eq:DB-th=}\\
  \left( \begin{aligned}
{    E}^{(-)}_\phi& \\
{    H}^{(-)}_\phi& \\
  \end{aligned} \right)_{\tt ob} &=&  ik\frac{e^{-ikr}}{r}\Big\{-\frac{\partial f^{(-)}_{\tt ob}(\theta)}{\partial \theta}+\frac{f^{(-)}_{\tt ob}(\theta)}{\sin\theta}\Big\}
\left( \begin{aligned}
\sin\phi \\
-\cos\phi  \\
  \end{aligned} \right) \,e^{-i\omega t}.
  \label{eq:DB-ph=}
\end{eqnarray}

Using (\ref{eq:Pi_g+p22=0=f}), we may rewrite the expression in curly braces as follows:
\begin{eqnarray}
-\frac{\partial f^{(-)}_{\tt ob}(\theta)}{\partial \theta}+\frac{f^{(-)}_{\tt ob}(\theta)}{\sin\theta}=\frac{E_0}{2k^2}\sum_{\ell=1}^{kR_0} (-1)^{\ell}\frac{2\ell+1}{\ell(\ell+1)}
e^{-i\frac{\ell(\ell+1)}{2kr}}
\Big\{-\frac{\partial P^{(1)}_\ell(\cos\theta) }{\partial \theta}+\frac{P^{(1)}_\ell(\cos\theta) }{\sin\theta}\Big\}.
  \label{eq:f_vp-}
\end{eqnarray}
{}
Introducing the amplitude function
\begin{eqnarray}
S^{(-)}(\theta)&=& \frac{1}{2}
\sum_{\ell=1}^{kR_0} (-1)^{\ell}\frac{2\ell+1}{\ell(\ell+1)}e^{-i\frac{\ell(\ell+1)}{2kr}}
\Big\{-\frac{\partial P^{(1)}_\ell(\cos\theta) }{\partial \theta}+\frac{P^{(1)}_\ell(\cos\theta) }{\sin\theta}
\Big\},
\label{eq:S1-v0p=}
\end{eqnarray}
we can present (\ref{eq:DB-th=})--(\ref{eq:DB-ph=}) as
{}
\begin{eqnarray}
  \left( \begin{aligned}
{    E}^{(-)}_\theta& \\
{    H}^{(-)}_\theta& \\
  \end{aligned} \right)_{\tt ob} &=& -E_0\left( \begin{aligned}
\cos\phi \\
\sin\phi  \\
  \end{aligned} \right)\frac{e^{-i(kr+\omega t)}}{ikr}S^{(-)}(\theta)
,\quad
\left( \begin{aligned}
{    E}^{(-)}_\phi& \\
{    H}^{(-)}_\phi& \\
  \end{aligned} \right)_{\tt ob} =E_0\left( \begin{aligned}
-\sin\phi \\
\cos\phi  \\
  \end{aligned} \right)\frac{e^{-i(kr+\omega t)}}{ikr}S^{(-)}(\theta).
  \label{eq:DB-t-pl=}
\end{eqnarray}

To evaluate  (\ref{eq:S1-v0p=}), we  use expressions (\ref{eq:pi-l})--(\ref{eq:tau-l}),  that result in
\begin{eqnarray}
S^{(-)}(\theta)&=&\sum_{\ell=1}^{kR_0} (-1)^{\ell}(\ell+{\textstyle\frac{1}{2}})e^{-i\frac{\ell(\ell+1)}{2kr}}
J_2\big((\ell+{\textstyle\frac{1}{2}})\theta\big).
\label{eq:S1-v7}
\end{eqnarray}
Taking $\theta=0$, we see that $S^{(-)}(\theta)=0$ and thus, the entire EM field (\ref{eq:DB-t-pl=}) also vanishes on the optical axis.

\section{Discussion and Conclusions}
\label{sec:end}

In this paper, we presented a flexible wave-theoretical description of the shadow cast by a large, fully absorbing sphere in the presence of a high frequency monochromatic plane incident wave. We presented the EM field in terms of Debye potentials. We then turned to Mie theory, utilizing a series expansion to represent the incident and scattered waves. We utilized the properties of the Riccati-Bessel functions to demonstrate that there is, indeed, complete cancelation in the shadow: the electromagnetic field vanishes in this region. 

Specifically, in Sec.~\ref{sec:Deb-pot} we have shown that the application of the fully absorbing boundary conditions introduced in Sec.~\ref{sec:bound-cond} leads to a complete geometric shadow behind the sphere. This shadow is given by $R_0=r\cos(\theta\mp\textstyle{\frac{\pi}{2}})$, which, in the chosen axially-symmetric coordinate system, represents a cylinder with a diameter set by the straight lines representing the rays of light that are just grazing the sphere. This is the geometric shadow for which the rays trajectories  (\ref{eq:theta-b0}) with the impact parameters within the range of $0\leq b\leq R_0$ are forbidden. On the other hand, the rays with impact parameters $b\gtrsim R_0+\lambda$  (from (\ref{eq:S-l-pri-p})), are fully transmitted towards the image plane.

In Sec.~\ref{sec:EM-field-near-axis} we explored the geometric shadow by looking for the EM field in the area close to the optical axis for $\theta\approx 0$ and  in Sec.~\ref{sec:EM-field-on-axis} we searched for light exactly on the axis where $\theta=0$. We were able to demonstrate that the results derived within the geometrical optics approximation and those achieved with the wave-optical treatment yield the same conclusion: the EM field is vanishing everywhere behind the sphere. Thus, there is no light in  the shadow. 

The exception is the EM field that is diffracted into the shadow by the edges of the obscuration. In accord to the classical diffraction theory (e.g., \cite{Born-Wolf:1999}), the corresponding light wanders inside the shadow with the usual  diffraction-limited divergence angle of $\sim \lambda/D_0$. For the SGL this ratio is $\lambda/D_0\ll1$; in addition, the solar boundary is very turbulent, so these effects were not discussed here. On the other hand, one could easily incorporate in the analysis effects related to the diffraction of light on the sharp edges of a sphere  by using the tools developed here. 

In addition, we note that the fully absorbing boundary conditions do not capture the interaction of light at the physical surface of the sphere. As a result, these conditions do not yield a description of the spot of Arago. Although this task is beyond the scope of the present paper, one may incorporate this feature into the wave-optical treatment  presented here by using, for instance, the approximation scheme introduced in \cite{Keller-etal:1956,Grandy-book-2000}. 

Finally, instead of the fully absorbing boundary conditions, one may choose a different set of boundary conditions associated with the sphere. For instance, the sphere may allow for some light transmission, it may have smooth surface and permit some re-emission, etc. By way of example, semitransparent boundary conditions discussed in \cite{Greider-Glassgold:1960} would result in a phase shift for the fictitious EM field introduced by the obscuration. To implement such boundary conditions, following the logic given by (\ref{eq:Pi_g+p22}), one would have to multiply the incoming wave (i.e., the wave that behaves as $\propto \zeta^{(+)}_\ell$) by the function that encapsulates the properties of the boundary and subtract the result from the incident EM field; once that is done, just follow the approach presented here to find the resultant EM field.  This approach may have some practical applications for detecting light behind an obstacle and will be explored further.

The wave-optical treatment presented here allows one to create a description of the scattering of light by a large sphere that can be readily extended to incorporate other effects such as those due to surrounding medium. Specifically, as we look forward to developing a full wave-theoretical description of the SGL \cite{Turyshev:2017,Turyshev-Toth:2017}, our objective is to develop an approach that can allow one to incorporate the gravitational effects on light produced by an extended body, as well as the effects of solar plasma. This work is on-going and the results, once available, will be published separately.

\begin{acknowledgments}
This work was performed at the Jet Propulsion Laboratory, California Institute of Technology, under a contract with the National Aeronautics and Space Administration.

\end{acknowledgments}


\appendix
\section{Riccati-Bessel functions}
\label{sec:Riccati-B}

The general solution to the radial equation (\ref{eq:R-bar-k*}) is well known and is given as a linear combination of the Riccati-Bessel functions $\psi_\ell(kr)$ and $\chi_\ell(kr)$  \cite{Born-Wolf:1999,Kerker-book:1969} as below:
{}
\begin{eqnarray}
R_\ell(kr)&=&c_\ell \psi_\ell(kr)+ d_\ell\chi_\ell(kr),
\label{eq:R-s}
\end{eqnarray}
where $c_\ell$ and $d_\ell$ are arbitrary constants. The functions $\psi_\ell(kr)$ and $\chi_\ell(kr)$ are related to the half integral order Bessel and Neumann functions $J_{\ell+{\textstyle\frac{1}{2}}}(kr)$ and $N_{\ell+{\textstyle\frac{1}{2}}}(kr)$ by
{}
\begin{eqnarray}
\psi_\ell(kr)=\sqrt{\frac{\pi kr}{2}}J_{\ell+{\textstyle\frac{1}{2}}}(kr), \qquad \qquad
\chi_\ell(kr)=-\sqrt{\frac{\pi kr}{2}}N_{\ell+{\textstyle\frac{1}{2}}}(kr).
\label{eq:R-s2}
\end{eqnarray}

Consider the asymptotic behavior of the Bessel and Neumann functions for large values of the argument, $kr\rightarrow\infty$. It is also well known and is given by \cite{Korn-Korn:1968} as
{}
\begin{eqnarray}
J_{\ell+\frac{1}{2}}(kr)&=&\sqrt\frac{2}{\pi kr}\Big(
{\cal A}_{\ell+\frac{1}{2}}(kr)\cos\big(kr-\frac{\pi}{2}(\ell+1)\big)-{\cal B}_{\ell+\frac{1}{2}}(kr)\sin\big(kr-\frac{\pi}{2}(\ell+1)\big)\Big), \label{eq:spher-Bess-New+}
\\
N_{\ell+\frac{1}{2}}(kr)&=&\sqrt\frac{2}{\pi kr}\Big(
{\cal A}_{l+\frac{1}{2}}(kr)\sin\big(kr-\frac{\pi}{2}(\ell+1)\big)+{\cal B}_{l+\frac{1}{2}}(kr)\cos\big(kr-\frac{\pi}{2}(\ell+1)\big)\Big),
\label{eq:spher-Newm-New+}
\end{eqnarray}
where the coefficients ${\cal A}_{\ell+\frac{1}{2}}$ and ${\cal B}_{\ell+\frac{1}{2}}$ have the following asymptotic behavior:
{}
\begin{eqnarray}
{\cal A}_{\ell+\frac{1}{2}}(kr)&\simeq&1-\frac{\ell(\ell+1)\big(\ell(\ell+1)-2\big)}{8(kr)^2}+\frac{\ell(\ell+1)\big(\ell(\ell+1)-2\big)\big(\ell(\ell+1)-6\big)\big(\ell(\ell+1)-12\big)}{384(kr)^4}+{\cal O}\big((kr)^{-6}\big),\qquad
\nonumber\\
{\cal B}_{\ell+\frac{1}{2}}(kr)&\simeq&\frac{\ell(\ell+1)}{2kr}-\frac{\ell(\ell+1)\big(\ell(\ell+1)-2\big)\big(\ell(\ell+1)-6\big)}{48(kr)^3}+{\cal O}\big((kr)^{-5}\big).
\label{eq:J-AB0}
\end{eqnarray}

Taking (\ref{eq:spher-Bess-New+})--(\ref{eq:J-AB0}) into account, we obtain the asymptotic behavior of $\psi_\ell(kr)$ and $\chi_\ell(kr)$ from  (\ref{eq:R-s2}) for large values of the argument when $kr\rightarrow\infty$:
{}
\begin{eqnarray}
\psi_\ell(kr)&=&\sqrt\frac{\pi kr}{2}J_{\ell+\frac{1}{2}}(kr)=
\cos\Big(kr-\frac{\pi}{2}(\ell+1)+\frac{\ell(\ell+1)}{2kr}\Big)+
{\cal O}\big((kr)^{-2}\big), \label{eq:spher-Bess-New+a}
\\
\chi_\ell(kr)&=&-\sqrt\frac{\pi kr}{2}N_{\ell+\frac{1}{2}}(kr)=
-\sin\Big(kr-\frac{\pi}{2}(\ell+1)+\frac{\ell(\ell+1)}{2kr}\Big)+{\cal O}\big((kr)^{-2}\big).
\label{eq:spher-Newm-New+a}
\end{eqnarray}
Note that a similar functional dependence, including the ${\ell(\ell+1)}/{2kr}$ term, was obtained in \cite{Turyshev-Toth:2017}, considering the solution to (\ref{eq:R-bar-k*}) using the WKB approximation and extending it closer to the turning point.

From expression (\ref{eq:R-s2}) we can form two linear combinations of  functions $\psi_l(k\rho)$ and $\chi_l(kr)$, namely
{}
\begin{eqnarray}
\zeta^{(+)}_\ell(kr)&=&\psi_\ell(k\rho)-i\chi_\ell(kr)=
\exp\Big[i\Big(kr-\frac{\pi}{2}(\ell+1)+\frac{\ell(\ell+1)}{2kr}\Big)\Big]+
{\cal O}\big((kr)^{-2}\big),
\label{eq:spher-Hank1}\\
\zeta^{(-)}_\ell(kr)&=&\psi_\ell(kr)+i\chi_\ell(kr)=
\exp\Big[-i\Big(kr-\frac{\pi}{2}(\ell+1)+\frac{\ell(\ell+1)}{2kr}\Big)\Big]+
{\cal O}\big((kr)^{-2}\big).
\label{eq:spher-Hank2}
\end{eqnarray}

For the chosen geometry, the two expressions above have a clear physical meaning \cite{Thomson-Nunes-book:2009}. They represent two waves: that given by (\ref{eq:spher-Hank1}), which is moving from the source toward and past the sphere and then on to the positive infinity,  called radially incoming wave; and that given by (\ref{eq:spher-Hank2}), which is moving in the opposite direction, called radially outgoing.

\end{document}